\documentclass{article}
\usepackage{epsfig}

\date{\today}
\begin{document}
\begin{center}
{\Large\bf Euclidean solutions in Einstein-Yang-Mills-dilaton theory}
\vspace{0.5cm}
\\
{\bf Yves Brihaye$^{1}$ } and~
{\bf Eugen Radu$^{2}$   }

\vspace*{0.2cm}
{\it $^1$Physique-Math\'ematique, Universite de
Mons-Hainaut, Mons, Belgium}

{\it $^2$ Department of
Mathematical Physics, National University of Ireland Maynooth, Maynooth,
Ireland}
\vspace{0.5cm}
\end{center}
\begin{abstract}  
We present arguments for the existence of a new type of 
solutions of the Euclidean
Einstein-Yang-Mills-dilaton theory in $d=4$ dimensions. 
Possesing nonvanishing nonabelian charges, these nonselfdual
configurations have no counterparts
on the Lorentzian section.
They provide, however, new saddle points in the Euclidean path integral.
\end{abstract}
%%%%%%%%%%%%%%%%%%%%%%%%%%%%%%%%%%%%%%%%%%%%%%%%%%%%%%%%%%%%%%%%%%%%
%%%%%%%%%%%%%%%%%%%%%%%%%%%%%%%%%%%%%%%%%%%%%%%%%%%%%%%%%%%%%%%%%%%%%%%%%%%%
\section{Introduction}
%%%%%%%%%%%%%%%%%%%%%%%%%%%%%%%%%%%%%%%%%%%%%%%%%%%%%%%%%%%%%%%%%%%%%%%%%%%%
Following the
discovery by Bartnik and McKinnon (BK) of
particle-like solutions of the four-dimensional
Einstein-Yang-Mills (EYM) equations
\cite{Bartnik:1988am}, there has been much interest
in classical solutions
of Einstein gravity  with nonabelian matter sources 
(for a review see \cite{Volkov:1999cc}).
These include hairy black holes solutions, which led to
a revision of some of the basic concepts of black hole physics
based on the uniqueness and no-hair theorems \cite{Volkov:sv}.

However, most of the investigations in the literature 
have been carried out for the physically most relevant case
of a spacetime with Lorentzian signature. 
The selfdual YM instantons  \cite{Charap:1977ww} 
although presenting interesting features,
have a vanishing energy momentum tensor and do not 
disturb the spacetime geometry.
The case of nonabelian gravitating instantons $i.e.$ nonselfdual solutions of the field equations 
with Euclidean metric signature, received only scarce attention.
Such configurations play an important role in the Euclidean approach to quantum gravity, 
where quantum amplitudes 
are defined by sum over positive definite metrics \cite{Gibbons:1976ue}.
In the saddle point approximation, these sums are dominated 
by gravitational instantons with appropriate boundary conditions.

In the Einstein-Maxwell theory, the Euclidean solutions are found
 analytically continuing the
Lorentzian signature configurations  \cite{Gibbons:1976ue, Hawking:1995ap}.
The situation is more complicated 
for a nonabelian matter content.
First, due to the absence of
closed form solutions, the field equations should be solved numerically.
However, for nonabelian configurations with  magnetic  fields only,
 the analytical continuation in time coordinate
has no effects at the level of the field equations which
 have the same form on both Lorentzian and 
Euclidean sections.
In particular, the  BK 
particle-like solutions (and the corresponding black hole generalizations)
extremize the EYM action also on the Euclidean section.
The Euclidean action of these solutions can straightforwardly be evaluated \cite{Moss:1992ne}; 
in particular
the black hole solutions satisfy the fist law of thermodynamics
and the standard $S=A_H/4$ entropy-area relation.

The situation is more complicated for  nonabelian 
solutions  possessing both electric
and magnetic nonabelian fields.
The crucial point there is that the asymptotic value of the electric
nonabelian potential $A_t$ has a direct physical relevance and cannot take arbitrary values.
Restricting to a nonanabelian SU(2) field,
one finds that the asymptotically
flat Lorentzian solutions necessarily have a vanishing electric
field $A_t=0$ and no nonabelian magnetic charge \cite{Ershov:1991nv, Bizon:1992pi}.
However, here we present both numerical and analytical arguments 
for the existence of nontrivial solutions with both electric 
and magnetic charges for
an Euclidean spacetime signature, within the same metric ansatz.
Possessing a finite mass and action, these nonselfdual
configurations have no Lorentzian counterparts
and provide new saddle points in the Euclidean path integral.
Both particle-like solutions with a $R^4$ topology and Euclidean black holes
with a topology $S^2\times R^2$ are found.
In all cases, the nonabelian fields are nonselfdual and curve the Euclidean geometry.
 
The paper is structured as follows:  in the next Section   we present the general
framework and analyse the field equations and boundary conditions.
In Section 3 we present our numerical results, while in
Section 4 we analyze  the Euclidean action 
and present a brief discussion of the thermodynamics
of solutions.
We conclude in Section 5
with a discussion of our results.

%%%%%%%%%%%%%%%%%%%%%%%%%%%%%%%%%%%%%%%%%%%%%%%%%%%%%%%%%%%%%%%%%%%%%%%%%%%%%%
%%%%%%            general action
%%%%%%%%%%%%%%%%%%%%%%%%%%%%%%%%%%%%%%%%%%%%%%%%%%%%%%%%%%%%%%%%%%%%%%%%%%%%%%
\section{General framework and equations of motion}
\subsection{Action principle }
 Our study of the Euclidean Einstein-Yang-Mills-dilaton
 (EYMd) system is based upon the following  generalization
 of the EYM action
\begin{eqnarray}
\label{lag0} 
I=-\int_{ \mathcal{M}}
d^4x \sqrt{g} \left ( \frac{R}{16\pi G}-\frac{1}{2}\partial_a \phi \partial^a \phi -
\frac{1}{2} e^{2\gamma \phi} {\rm Tr} (F_{ab} F^{ab}) 
   \right ) 
  -\frac{1}{8\pi G}\int_{\partial\mathcal{M}} d^{3}x\sqrt{h}K.
\end{eqnarray}
This type of action  
arises from various unified theories including superstring
models and, apart from gravity and YM fields,
contains also a dilaton field $\phi$ with coupling constant $\gamma$.
(For example, $\gamma=1/2$ and $\gamma=1/\sqrt{3}$ are 
the cases corresponding to string theory and Kaluza-Klein reduction
of the $d=5$ EYM theory, respectively.
The usual EYM system is found for $\phi=0$, $\gamma=0$.)
Since changing the sign of $\gamma$ is equivalent to changing the sign of 
$\phi$, it is sufficient to consider only $\gamma>0$.

The field strength
tensor is given by 
$F_{ab}=\partial_{a}A_{b}-\partial_{b}A_{a}
-ig[A_{a},A_{b}],$
with 
$g$ the Yang-Mills  coupling constant.
The last term in  (\ref{lag0}) is the Hawking-Gibbons surface term \cite{Gibbons:1976ue}, 
where $K$ is the trace 
of the extrinsic curvature for the boundary $\partial\mathcal{M}$ and $h$ is the induced 
metric of the boundary.

Asking for stationary points of this action, one finds 
the EYMd field equations
 \begin{eqnarray}
\label{Einstein-eqs}
R_{ab}-\frac{1}{2}g_{ab}R =
8\pi G  T_{ab}~,~~~
\nabla^2 \phi-\gamma e^{2\gamma \phi} Tr(F_{ab}F^{ab})=0~,~~~
\nabla_{a}(e^{2 \gamma \phi}F^{ab})-ige^{2\gamma \phi}[A_{a},F^{ab}]=0~,
\end{eqnarray}
where the energy-momentum tensor is defined by
\begin{eqnarray}
\label{Tij}
T_{ab} =
\partial_{a} \phi \partial_{b} \phi
-\frac{1}{2}g_{ab} \partial_{c}\phi \partial^c \phi +
2e^{2\gamma \phi}{\rm Tr}
    ( F_{ac} F_{bd} g^{cd}
   -\frac{1}{4} g_{ab} F_{cd} F^{cd}).
\end{eqnarray}

%%%%%%%%%%%%%%%%%%%%%%%%%%%%%%%%%%%%%%%%%%%%%%%%%%%%%%%%
%\section{Spherically symmetric solutions}
%%%%%%%%%%%%%%%%%%%%%%%%%%%%%%%%%%%%%%%%%%%%%%%%%%%%%%%%
%%%%%%%%%%%%%%%%%%%%%%%%%%%%%%%%%%%%%%%%%%%%%%%%%%%%%%%%%%%%%%%%%%%%%
\subsection{The ansatz}
%%%%%%%%%%%%%%%%%%%%%%%%%%%%%%%%%%%%%%%%%%%%%%%%%%%%%%%%%%%%%%%%%%%%%
We consider the following spherically symmetric metric with Euclidean signature
\begin{eqnarray}
\label{MSD}
ds^2=  
\sigma^2(r) N(r) d\tau^2 + \frac{dr^2}{N(r)} 
+ r^2(d \theta^2 + \sin^2 \theta d \varphi^2),
\end{eqnarray}
where $r$ is the radial coordinate, $\theta$ and $\varphi$ are the angular
coordinates with the usual range, while $\tau$ corresponds to the Euclidean time.

In this paper we'll discuss two types of solutions. 
The first type of configurations, which usually corresponds to analytical
continuations of Lorentzian globally regular, particle-like solutions
have $0\leq r<\infty$ and a trivial topology,
the Killing vector 
$\partial/\partial \tau$ presenting no fixed points sets 
($i.e.$ $g_{\tau \tau}>0$ for any $r$).
For the second type of solutions, the fixed
point set of the Euclidean time
symmetry  is of two dimensions (a "bolt") and 
the range of the radial coordinate is restricted
to $0<r_h\leq r<\infty$, with $N(r_h)=0$ and a finite nonzero $\sigma(r_h)$.
 
For the first type of solutions, 
the periodicity $\beta$ of the Euclidean time coordinate is arbitrary,
while for bolt solutions $\beta$ is fixed by regularity requirements to
\begin{eqnarray}
\label{beta}
\beta=\frac{4 \pi}{\sigma(r_h) N'(r_h)}~,
\end{eqnarray}
where a prime denotes the derivative with respect to $r$.

Without any loss of generality, the  SU(2) ansatz
can be written in the form
\begin{eqnarray}
\label{YMansatz}
A_a dx^a=\frac{1}{2g} \{ u(r) \tau_3 d\tau+  
 w(r) \tau_1  d \theta
+\left(\cot \theta \tau_3
+ w(r) \tau_2 \right) \sin \theta d \varphi \}, 
\end{eqnarray}
(where the  
$\tau_i$ are the standard Pauli matrices),
being described by two 
functions $w(r)$ and $u(r)$ which we shall refer to as magnetic 
and electric potential, respectively.

Therefore the spherically symmetric EYMd system is described
by the following one-dimensional effective action
\begin{eqnarray}
{\cal L} = \frac{1}{4\pi G} \sigma m'  
-  \bigg[ \frac{e^{2 \gamma \phi}}{g^2}(
\sigma N  w'^2  
+\frac{1}{2r^2} \sigma(1-\omega^2 )^2
~~
+ \frac{1}{2\sigma} r^2 u'^2 
+ \frac{1}{\sigma N}  \omega^2   u^2)
 -\frac{1}{2} \sigma N r^2 (\phi')^2,
 \bigg]
\end{eqnarray}
%%%%%%%%%%%%%%%%%%%%%%%%%%%%%%%%%%%%%%%%%%%%%%%%%%%%%%%%%%%%%%%%%
the field  equations  reducing to a set of four
non-linear differential equations
\begin{eqnarray}
\nonumber
m' = 4\pi G\bigg[ \frac{e^{2 \gamma \phi}}{g^2}
    \bigg( N w'^2 + \frac{(1-w^2)^2}{2 r^2} 
      - \frac{1}{2\sigma^2} r^2 u'^2 - \frac{1}{\sigma^2N} w^2 u^2\bigg)
     +    \frac{1}{2} N r^2 \phi'^2\bigg],
\\
\label{e2}
\sigma' = \frac{8\pi G\sigma}{r} 
(\frac{1}{2} N r^2 \phi'^2  
+ 2 \frac{e^{2 \gamma \phi}}{g^2} ( w'^2 - \frac{1}{\sigma^2 N^2} w^2 u^2) ),~~~~
%\\
%\label{e3}
(\sigma N e^{2 \gamma \phi} w')'=
e^{2 \gamma \phi}(\frac{1}{r^2}\sigma w(w^2-1) + \frac{w u^2}{\sigma N} ),
\\
\nonumber
(e^{2 \gamma \phi} \frac{r^2 u'}{\sigma})' = 
\frac{2 w^2 u}{\sigma N} e^{2 \gamma \phi},~~~~
%\\
%\label{e5}
(\sigma N r^2 \phi')' = \frac{2 \gamma e^{2 \gamma \phi}}{g^2}
\bigg(\sigma N w'^2 + \frac{\sigma(1-w^2)^2}{2 r^2} 
+ \frac{r^2 u'^2}{2 \sigma} + \frac{w^2 u^2}{\sigma N} \bigg).
\end{eqnarray}
We are interested in asymptotically flat
regular solutions whose mass function $m(r)$ approaches a constant finite value
as $r \to \infty$, which will have also a finite Euclidean action.

Solutions of the above system
are already known in a few particular cases. 
The embedded U(1) configurations correspond to \cite{gibbons, horowitz}
(in units $4 \pi G=g=1$)
\begin{eqnarray}
\label{m2}
ds^2=\frac{dr^2}{\lambda^2}+
R^2(d \theta^2 + \sin^2 \theta d \varphi^2)+\lambda^2 d\tau^2,
~~\lambda^2=(1-\frac{r_{+}}{r})(1-\frac{r_{-}}{r})^{(1-\gamma^2)/(1+\gamma^2)},
\\
\nonumber
R=r(1-\frac{r_{-}}{r})^{ \gamma^2/(1+\gamma^2)},~~
e^{2\phi}=(1-\frac{r_{-}}{r})^{2\gamma/(1+\gamma^2)},~~~w=0, 
\end{eqnarray}
with an arbitrary value of the electric potential $u(r)=\Phi$, $r_+$ being a positive constant and
$r_{-}=(1+\gamma^2)/r_+$.

In the case $\gamma=0$ one can set 
$\phi=0$ and we find the selfdual YM dyonic solution
 $w=r/\sinh r$, $ u=\coth r-1/r$ in a flat spacetime background.
 For bolt configurations with  $N=1-2M/r$, $\sigma\equiv 1$, the analogous of this solution
solves the equations
\begin{eqnarray}
\label{newSD}
w'=wu/N,~~~r^2 u'=w^2-1,
\end{eqnarray}  
and corresponds to a selfdual dyonic instanton in a Schwarzschild
 background.
Except for the region near $r=r_h$, the numerical
solution of these equations
looks very similar to the flat space counterpart.

For a truncation $A_\tau=0$ of the YM ansatz,
the purely magnetic Euclidean configurations are found by taking $t \to i\tau$
in the Lorentzian solutions describing EYMd particle-like and black hole 
solutions discussed in \cite{Lavrelashvili:1992ia,Donets:1992zb}. 
Within a stationary ansatz, this analytical continuation 
has no effect at the level of equations
of motion.
Features of the corresponding EYM Euclidean solutions 
have been discussed in \cite{Moss:1992ne}.
For $u(r)=0$ and $\gamma=1/2$, the lowest energy solution corresponds to a
BPS configuration in a version of ${\cal{N}}=4$, $D=4$ gauged supergravity
\cite{Volkov:1999vu}.
Excited configurations indexed by the node number of $w$ exist as well.
However, all these solutions have zero magnetic charge.

In fact, for any $\gamma$, the only spherically symmetric 
Lorentzian configurations with reasonable asymptotics are the purely magnetic
EYMd particle-like and black hole 
solutions \cite{Lavrelashvili:1992ia,Donets:1992zb}.
For finite mass solutions,
the electric potential should vanish identically,
the arguments presented in \cite{Ershov:1991nv, Bizon:1992pi}
being not affected by the presence of a dilaton.
First, one can see from eqs. (\ref{e2}) that for $A_\tau\neq 0$, 
the magnitude at infinity of the electric 
potential should be nonzero.
However, for Lorentzian signature, this 
would imply from the $w-$equation an oscilatory asymptotic behavior of the 
magnetic gauge function $w$ and thus an infinite mass. 
Therefore $u(\infty)=0$, which means 
a purely magnetic solution $u(r)\equiv0$ and $w(\infty)=\pm 1$, $i.e.$ no magnetic charge.

However, the situation is different for an Euclidean spacetime signature.
In this case, the electric potential qualitatively behaves as a Higgs field and nontrivial
electrically charged solutions may exist 
\footnote{Note that the electric potential-Higgs field analogy
is not complete, since the $A_\tau$
couples also to the $g_{\tau \tau}$ metric component.}.
Thus we expect these solutions to present
a magnetic charge and to share a number of common features with the usual monopoles.

%%%%%%%%%%%%%%%%%%%%% FIgure 1 %%%%%%%%%%%%%%%%%%%%%%%%%%%
\newpage
\setlength{\unitlength}{1cm}

\begin{picture}(18,7)
\label{fig1}
\centering
\put(2,-3.5){\epsfig{file=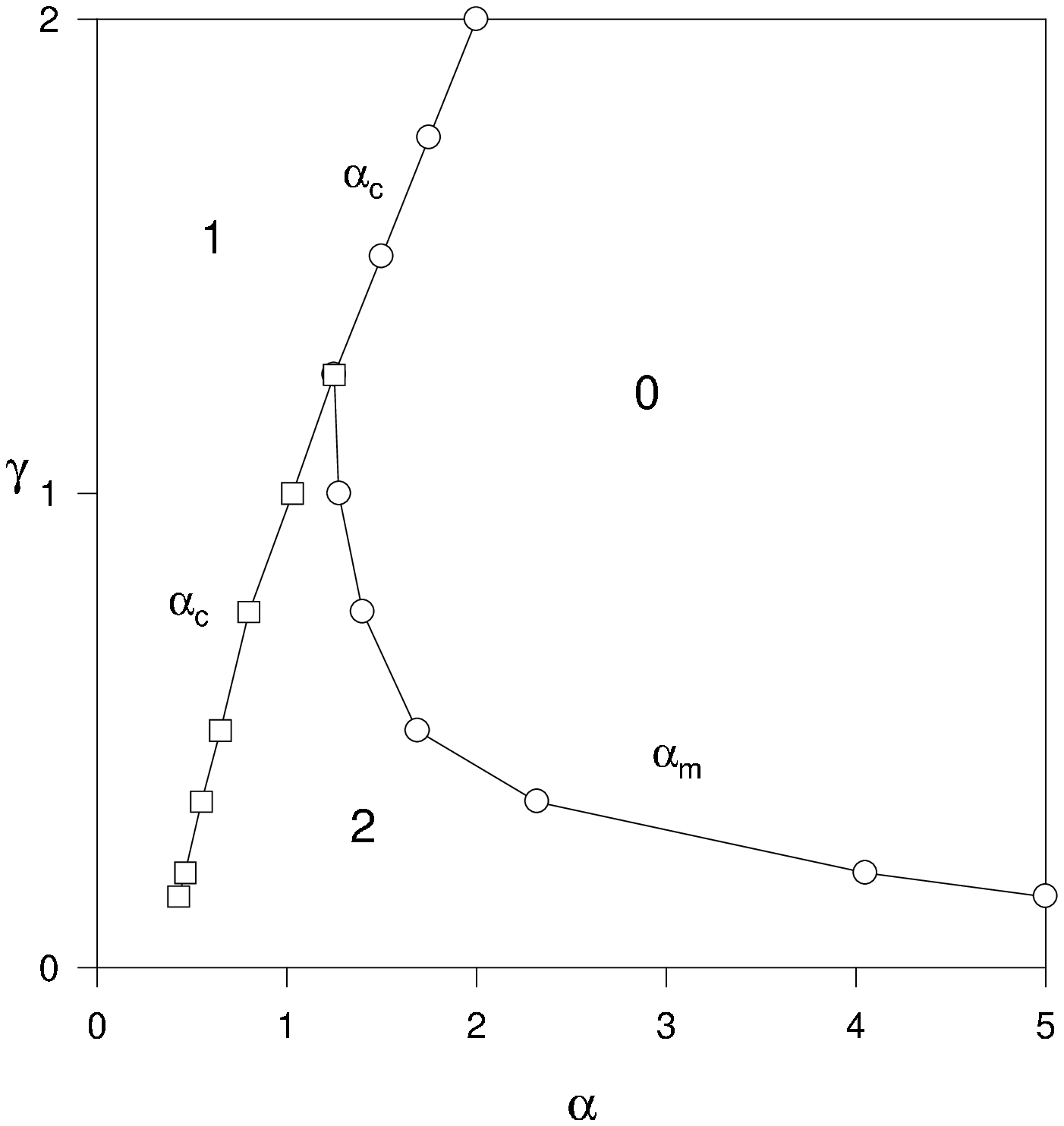,width=11cm}}
\end{picture} 
\\
\\
{\small {\bf Figure 1.} The domain of existence
in the $(\alpha, \gamma)$-plane of the non-abelian solutions
with $R^4$ topology. The labels $0,~1,~2$ 
refer to the number of solutions we found.}
\\
%%%%%%%%%%%%%%%%%%%%%%%%%%%%%%%%%%%%%%%%%%%%%%%%%%%%%%%%%%%%%%%%%%%%%
\subsection{Boundary conditions}
%%%%%%%%%%%%%%%%%%%%%%%%%%%%%%%%%%%%%%%%%%%%%%%%%%%%%%%%%%%%%%%%%%%%%
The field equations imply
the following behaviour for $r \to 0$ in terms of four parameters
$(b,\phi_0,\sigma_0,u_1)$:
\begin{eqnarray}
\label{bc1}
m(r)=8\pi G \frac{e^{2 \gamma \phi_0}}{g^2}(b^2-\frac{u_1}{4\sigma_0^2})r^3+O(r^4),
~~~
\sigma(r)=\sigma_0(1+4\pi G \frac{e^{2 \gamma \phi_0}}{g^2}(b^2-\frac{u_1}{4\sigma_0^2})r^2)
+O(r^4),
\\
\nonumber
w(r)=1-br^2+O(r^4),~~~u(r)=u_1r+O(r^3),
~~~
\phi(r)=\phi_0+2\gamma \frac{e^{2 \gamma \phi_0}}{g^2}(b^2+\frac{u_1}{4\sigma_0^2})r^2+O(r^4).
\end{eqnarray}
The corresponding expansion as $r \to r_h$ for bolt solutions is
\begin{eqnarray}
\label{eh} 
\nonumber 
&m(r)=r_h/2+m_1(r-r_h)+O(r-r_h)^2,
~~ 
\sigma(r)=\sigma_h+\sigma_1(r-r_h)+O(r-r_h)^2,
\\
&w(r)=w_h+\omega_1(r-r_h)+O(r-r_h)^2,
~~~
u(r)=u_1(r-r_h)+u_2(r-r_h)^2+O(r-r_h)^3,
\\
\nonumber 
&\phi(r)=\phi_h+\phi_1(r-r_h)+O(r-r_h)^2,
\end{eqnarray}
where
\begin{eqnarray}
\label{eh1} 
m_1=4\pi G \frac{e^{2 \gamma \phi_h}}{g^2}  (\frac{(1-w_h^2)^2}{2r_h^2}-\frac{r_h^2u_1^2}{2\sigma_h^2}),
~~~\sigma_1=\frac{4\pi G \sigma_h}{r_h}\bigg(\frac{e^{2 \gamma \phi_h}}{g^2}(w_1^2-
\frac{w_h^2 u_1^2}{N_1^2\sigma_h^2})+r_h^2\phi_1^2\bigg),
\\
w_1=\frac{w_h(w_h^2-1)}{N_1r_h^2},~~u_2=\frac{w_h^2 u_1}{r_h^2 N_1},
~~~
\phi_1=\frac{2\gamma e^{2\gamma\phi_h}}{\sigma_h r_h^2 N_1 g^2}
(\frac{\sigma_h}{2r_h^2}(1-w_h^2)^2+\frac{r_h^2u_1^2}{2\sigma_h}),
~~N_1=(1-2m_1)/r_h.
\end{eqnarray}
$w_h,\sigma_h,u_1,\phi_h$ being free parameters.

The analysis of the field equations as $r\to\infty$ gives the
following expression in terms of the constants $M,~\Phi,~\sigma_2,~Q_d,~Q_e$
\begin{eqnarray}
\label{bc3}
 N(r)=1-\frac{2M}{r}+..,~~\sigma(r)=1-\frac{2\pi G \sigma_2}{r^2}+..,~~
u(r)=\Phi-\frac{Q_e}{r}+..,~~
\phi(r)=-\frac{ Q_d }{r}+..,~~w(r)=e^{-\Phi r}+..,
\end{eqnarray} 
which is shared by both classes of solutions.
The physical significance of the quantity $\Phi$ is that gives the electrostatic
potential difference between the origin and infinity, while $M$ corresponds to the mass
of

%%%%%%%%%%%%%%%%%%%%% Figure 1 %%%%%%%%%%%%%%%%%%%%%%%%%%%
\newpage
\setlength{\unitlength}{1cm}

\begin{picture}(18,7)
\centering
\put(2,0.0){\epsfig{file=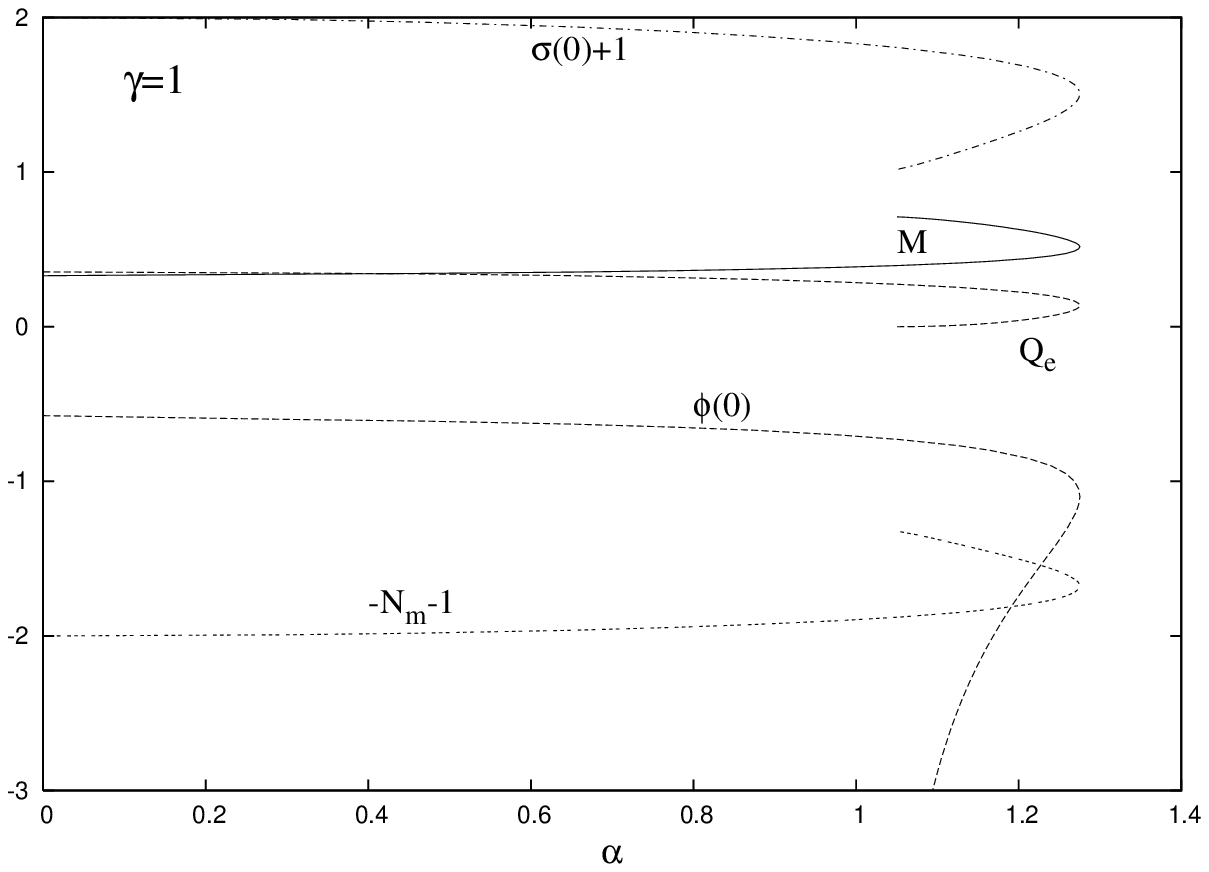,width=11cm}}
\end{picture}
\begin{picture}(19,8.)
\centering 
\put(2.4,0.0){\epsfig{file=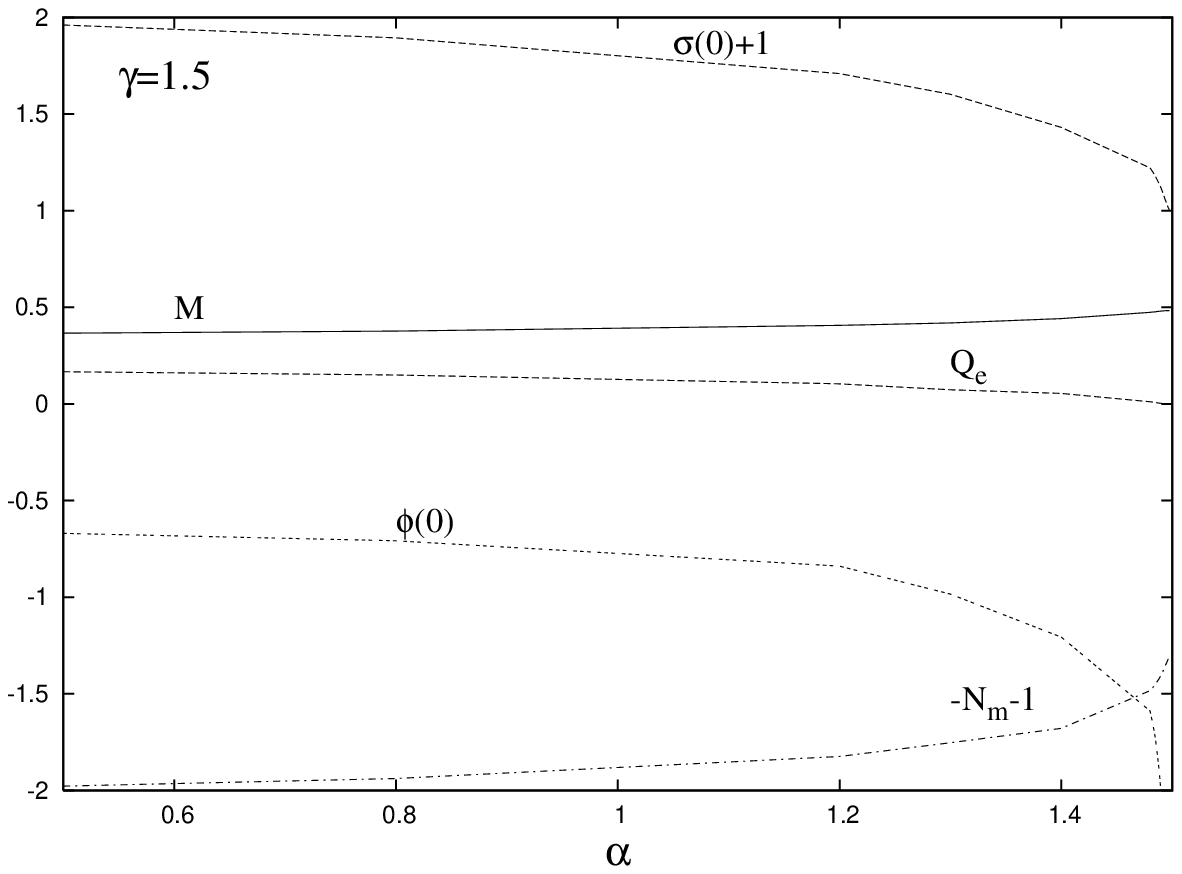,width=11.1cm}}
\end{picture}
\\
\\
{\small {\bf Figure 2.} The mass-parameter $M$, the nonabelian electric charge $Q_e$,
the values of the metric function
$\sigma(r)$ and dilaton function $\phi(r)$ at the origin
as well as the minimal value $N_m$ of the metric function $N(r)$  
 are represented as function of the parameter $\alpha$ for two values of $\gamma$.
 Here we consider solutions with $R^4$ topology.}
\\
\\
configurations.
These solutions possess also  nonabelian electric and magnetic charges
computed $e.g.$ according to
\begin{equation}
\label{def-charge}
\pmatrix{\bf{Q_E}\cr \bf{Q_M}\cr}
 = {1\over 4\pi} \int dS^{(k)}_{\mu} \, \sqrt{g} \,
{\rm Tr} \Big\{ 
 \pmatrix{ F^{\mu \tau}\cr \tilde F^{\mu \tau} \cr} \tau_3 
 \Big\}, 
\end{equation}
which are conserved from the Gauss flux theorem.
Thus the solutions with the asymptotic form ({\ref{bc3}) have 
a unit magnetic charge and a electric charge $Q_e$.
As discussed in Section 4, the dilaton charge $Q_d$ 
which enters the asymptotics (\ref{bc3}) is  fixed by other relevant
quantities.

%%%%%%%%%%%%%%%%%%%%% Figure 3 %%%%%%%%%%%%%%%%%%%%%%%%%%%
\newpage
\setlength{\unitlength}{1cm}

\begin{picture}(18,7)
\centering
\put(1.3,-1){\epsfig{file=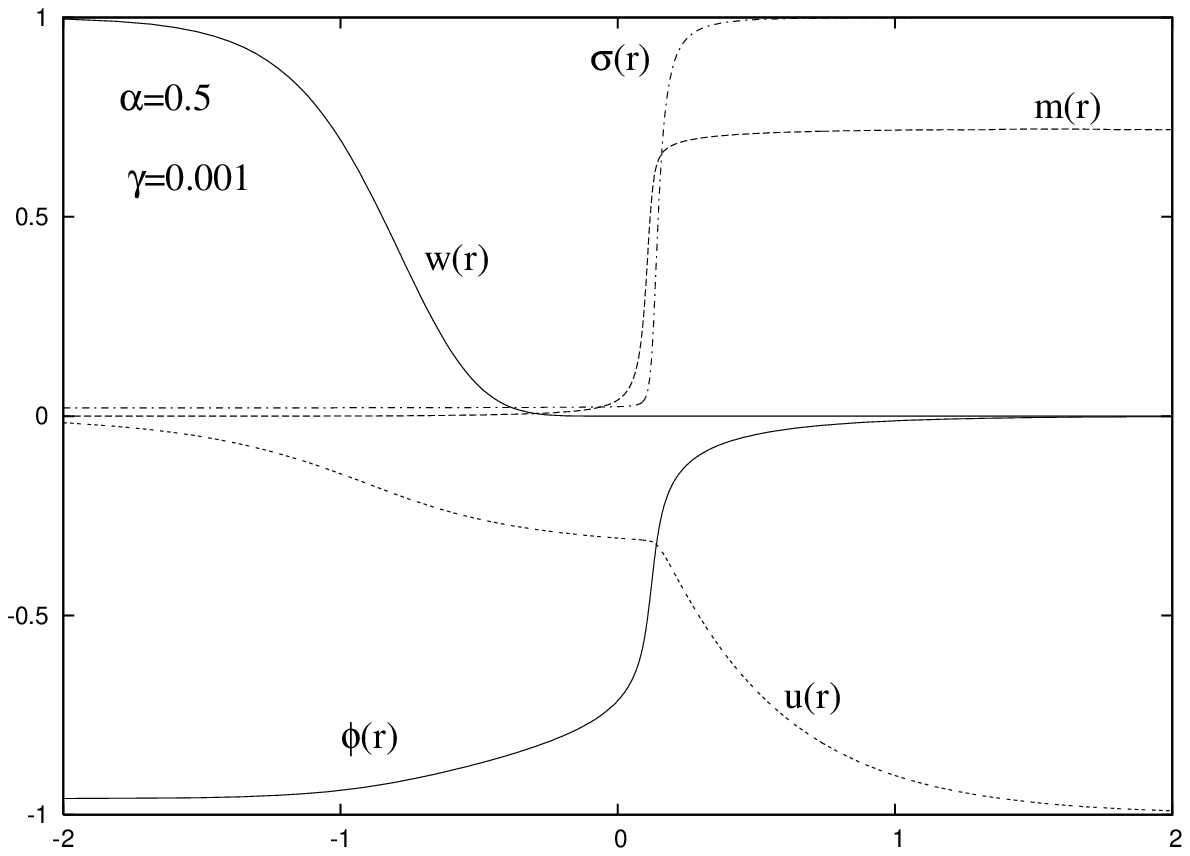,width=11cm}}
\end{picture} 
\\
\\
\\
\\
{\small {\bf Figure 3.} The profiles of a topologically $R^4$
EYMd solution are plotted for $\alpha=0.5$, $\gamma=0.001$.}
%%%%%%%%%%%%%%%%%%%%% FIgure 4 %%%%%%%%%%%%%%%%%%%%%%%%%%%
\\
\\
\\
\setlength{\unitlength}{1cm}
\begin{picture}(18,7)
\centering
\put(2,-1){\epsfig{file=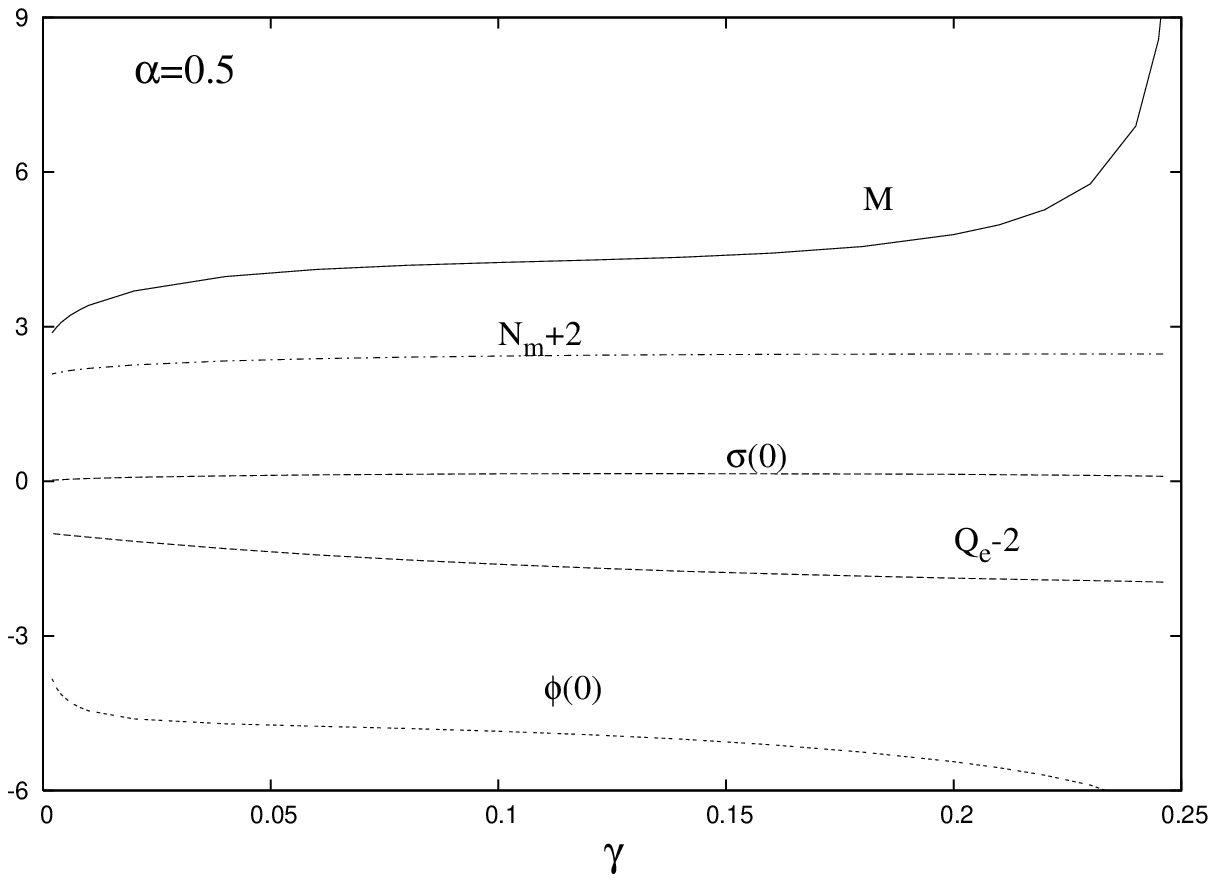,width=11cm}}
\end{picture} 
\\
\\
\\
\\
{\small {\bf Figure 4.} 
The same as Figure 2 for a fixed value of $\alpha$ and a varying dilaton
coupling constant $\gamma$.}
\\
%%%%%%%%%%%%%%%%%%%%%%%%%%%%%%%%%%%%%%%%%%%%%%%%%%%%%%%%%%
 \section{Numerical solutions}
%%%%%%%%%%%%%%%%%%%%%%%%%%%%%%%%%%%%%%%%%%%%%%%%%%%%%%%%%%
Although an analytic or approximate solution appears to be
intractable, here we present   arguments 
for the existence of  Euclidean solutions
satisfying the boundary conditions displayed above.
 
To perform numerical computations and order-of-magnitude estimations,
it is useful to have a new set of dimensionless variables.
This is obtained by using the following rescaling
$r\to r/(\Phi g)$, $m\to \Phi g m$, 
%%%%%%%%%%%%%%%%%%%%%%%%%%%%%%%%%%%%%%%%%%%%%%%%
\newpage
\setlength{\unitlength}{1cm}

\begin{picture}(18,7)
\centering
\put(2,0.0){\epsfig{file=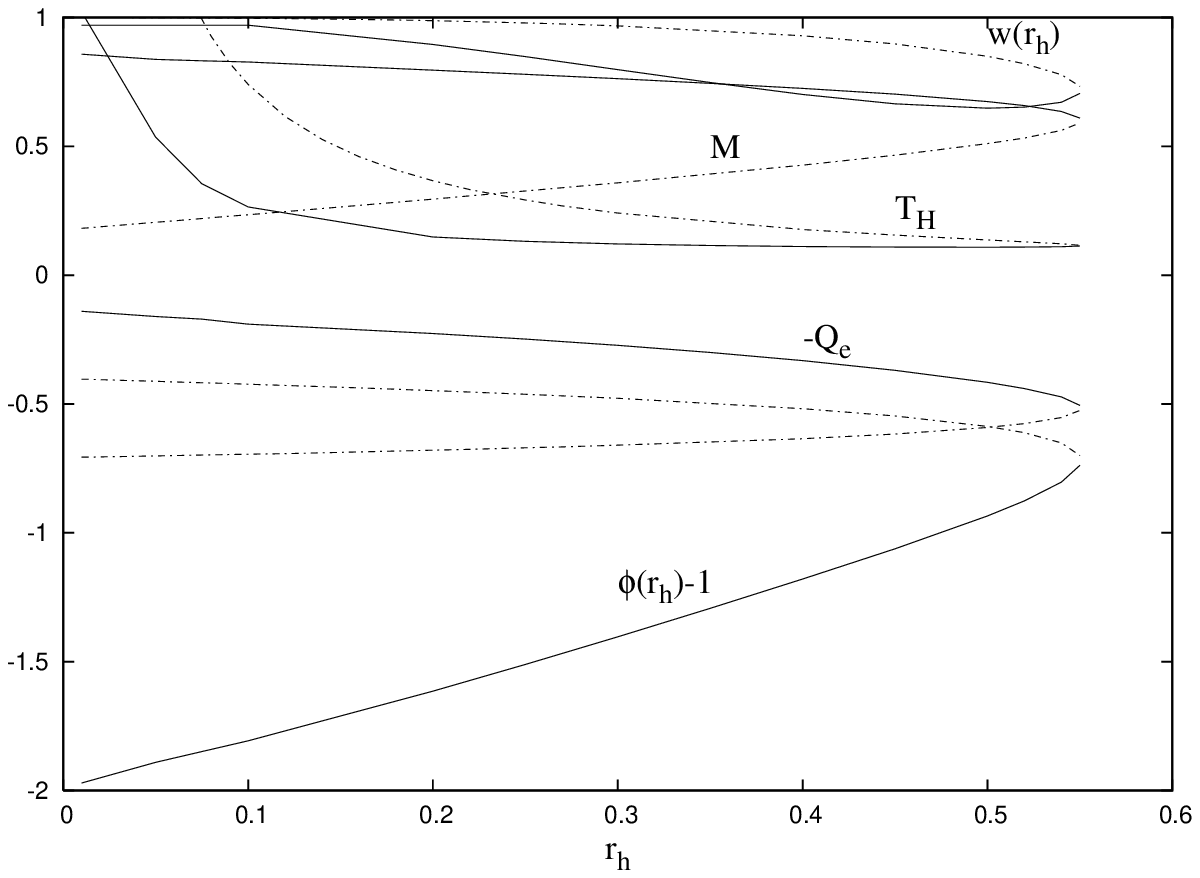,width=11cm}}
\end{picture} 
\\
\\
{\small {\bf Figure 5.} The mass-parameter $M$, the nonablelian electric charge $Q_e$,
the values of the metric function
$\sigma(r)$, the magnetic potential $w(r)$ and dilaton 
function $\phi(r)$ for $r=r_h$ 
 are represented as function of  $r_h$ for $\gamma = 0.5$, $\alpha = 1.0$
 bolt solutions.}
\\
\\
$ u \to u/(\Phi g)$,
where $\Phi$ is the
 asymptotic magnitude of the electric potential.
We consider also a  rescaling of the dilaton and  dilaton coupling constant 
$\phi \to \phi/\Phi$, $\gamma \to \Phi \gamma$. 
As a result, the field equations depend only on the
coupling constants $\alpha  = \sqrt{4\pi G}\Phi$ and the rescaled dilaton
constant $\gamma$.
%%%%%%%%%%%%%%%%%%%%%%%%%%%%%%%%%%%%%%%%%%%%%%%%%%%%%%%
\subsection{Topologically $R^4$ solutions}
%%%%%%%%%%%%%%%%%%%%%%%%%%%%%%%%%%%%%%%%%%%%%%%%%%%%%%%
The equations of motion (\ref{e2}) have been solved for a large set of
the parameters ($\alpha,~\gamma$), 
looking for solutions interpolating between the asymptotics (\ref{bc1}) 
and (\ref{bc3}).
For all solutions we
studied, the gauge functions $w(r)$ and $u(r)$, the 
metric functions $m(r)$, $\sigma(r)$ and the dilaton $\phi(r)$
interpolate monotonically 
between the corresponding values at $r=0$ and the asymptotic values at infinity, 
without presenting any local extrema.
Note also that for both
topologically $R^4$ and $R^2 \times S^2$ configurations, 
the asymptotic value of $m(r)$ was found to take positive values only.

The domain of existence of the solutions in the
 $(\alpha, \gamma)$-plane is rather involved.
According to the case $\gamma < \gamma_c$ or 
$\gamma > \gamma_c$ (with $\gamma_c \approx 1.25$) 
the pattern of the solutions is
quite different, as indicated on Figure 1, where the numbers in the
various domains of the graphic refers to the number of non-abelian
solutions in the corresponding domain we found.

First we discuss the  case $\gamma < \gamma_c$.
If we fix $\gamma$ in this range and increase $\alpha$ we
construct a branch of 
solutions which exists up to a maximal
value $\alpha_m(\gamma)$. 
Different quantities characterizing 
the solutions are reported on Figure 2 for $\gamma = 1$.
Then, our numerical analysis indicates that
a second branch of solutions exists 
for $\alpha \in [\alpha_c, \alpha_m]$ 
\footnote{Although we cannot exclude the existence of 
other secondary branches,
these would have a very small extension in $\alpha$. 
 These solutions are likely to exist at least for values of $\gamma$ around
$1/\sqrt{3}$ (see the remarks in Section 5).
Finding them is a difficult numerical problem, 
which is not aimed in this report.}.
The solutions of the first and of the second branch coincide in the
limit $\alpha \rightarrow \alpha_m$. In the limit 
$\alpha \rightarrow \alpha_c$, the solutions of the second branch
tends to an Einstein-Maxwell-dilaton configuration 
(\ref{m2}).  This is
characterized
 in the limit $\alpha \to \alpha_c$ ($\sim 1.034$ in the case $\gamma = 1.0$)
by the fact that $\phi(0)\to \infty$,
 $Q_e \to 0$ , $\sigma(r)\to 1$, $w(r)\to 0$.
 A few of these properties are visible on Figure 2 but the 
 claim is further demonstrated by comparing the profiles.

%%%Of course further evidence o seen by comparing the profiles
%%%of the two solutions.

It is natural to try to understand how these two branches of solutions
 behave for $\alpha$ fixed and for
$\gamma$ decreasing to zero.
Clearly, the solutions of the first branch tends to the flat space selfdual YM
solution mentioned in the previous section,
 $i.e.$ with  $\phi(r) = 0$, $N(r)=\sigma(r)=1$, $M=0$,
$Q_e=1$ and the functions $w(r),u(r)$ given by the PS dyon.
The numerical study of the limiting solution
(i.e. for $\gamma \rightarrow 0$) of the second branch strongly suggests
that it ends up into a discontinuous configuration indicated on Figure 3
for $\gamma = 0.001$. We see in particular that the dilaton function
$\phi$ and the metric function $\sigma$
presents a 
pronounced step at an intermediate value,  say $r=r_i$ of the radial
variable. At the same value of $r$, the metric function $N$
develops a deep  minimum. This critical phenomenon becomes more
severe while $\gamma$ deceases and the numerical integration
 strongly suggests that another solution of Eq.
(\ref{newSD}) is approached
on the interval $[r_i, \infty]$:
namely $\sigma(r)=1,w(r)=0, \phi(r)=0, u(r) = -1+1/r$ and $m$ is constant.
The counterpart of this result for Lorentzian case
 \cite{BFM,BHKT} is that gravitating monopoles and dyons form two branches
 of solutions, the second of which bifurcates into a Reissner-Nordstr\"om
 solution.

}

From Figure 1 we see that the solutions of the second branch exist
on a small interval of $\gamma$. The evolution of several parameters
characterizing the solutions of the second branch
are  reported on Figure 4 as functions of $\gamma$.

In the case $\gamma > \gamma_c$, the pattern is simpler, only one
solution is available and the branch of non-abelian solutions
bifurcate directly into a EMD solution for 
$\alpha \rightarrow \alpha_c$. This is illustrated by Figure 2 for $\gamma=1.5$.

%%%%%%%%%%%%%%%%%%%%%%%%%%%%%%%%%%%%%%%%%%%%%%%%%%%%%%%
\subsection{Solutions with $R^2 \times S^2$ topology}
%%%%%%%%%%%%%%%%%%%%%%%%%%%%%%%%%%%%%%%%%%%%%%%%%%%%%%%
 According to the standard arguments, one can expect
black hole generalisations of the configurations  discussed above to exist at
least for small values $r_h$.
This is confirmed by our numerical analysis and 
black hole solutions seem to exist for all values of $\alpha$
for which solutions with $R^4$ topology were constructed.

The first step is to determine the domain of existence of
solutions in the parameter space determined
by $(\alpha,\gamma, r_h)$. Since this constitutes a considerable
task, we limit here our numerical study to the case $\gamma=0.5$,
although solutions with other values of the dilaton 
coupling constant have been studied as well.
It is useful to notice that, corresponding to this value,
we have $\alpha_{m}\simeq 1.6$ and $\alpha_{c}\simeq 0.8$.

It turns out that black holes solutions exist in a finite domain
of the $(\alpha, r_h)$-plane. In other words, if we fix $\alpha$,
we are able to construct a first family
 of solutions up to a maximal value $r_{h,max}$
which depends on $\alpha$. As an example for $\alpha = 0.1$
and $\alpha = 1.0$ we find respectively $r_{h,max} \approx 0.75$
and  $r_{h,max} \approx 0.55$. In the limit $r_{h} \rightarrow 0$
the solutions on this branch,  characterized by the dashed 
lines in Figure 5,
approach the corresponding regular
solution on $r \in ]0,\infty[$.

However, the pattern of these black holes solutions for large values of $r_h$ 
is
more sophisticated and seems to be intimately related to the
number of topologically trivial solutions available for the corresponding
value of $\alpha$. Two cases  can be distinguished, 
according to $\alpha < \alpha_c$ and $\alpha_c < \alpha < \alpha_m$
for which respectively one and two $R^4$ solutions exist (see 
previous section).

  For $\alpha < \alpha_c$ the branch of black hole solutions
exists up to a maximal value of $x_{h,max}$. The corresponding value
$w(r_h)$ decreases and the solutions ends up into a configuration
with $w(r)\equiv 0$ when the value of $r_{h,max}$ is reached.

For $\alpha_c < \alpha < \alpha_m$,  a second branch of black hole
solutions exist.
 In the limit $r_h \rightarrow r_{h,max}$
the two solutions converge to a common solution. In the
limit $r_h \rightarrow 0$ the two branches approach the two
$R^4$ solutions available for this value of $\alpha$.
This is demonstrated on Figure 5 where the mass, the electric
charge, the temperature $T=1/\beta$
and the values $w(r_h)$, $\phi(r_h)$ are plotted as functions of $r_h$ for
$\gamma = 0.5$ and $\alpha = 1.0$;  there dashed (resp. dotted)
lines are used for the solutions of the first (resp. second) branch.
  
We also studied the evolution of the solutions in the
$\gamma \to 0$ limit. For the solutions of the main branch
we observe that the self-dual solution
(\ref{newSD}) is approached. Namely, the field of the dilaton
uniformly approaches $\phi(r)=0$. 
The action of the selfdual dyon (\ref{newSD}) is $4\pi \beta/g^2$ which,
 different from Charap-Duff solution  \cite{Charap:1977ww} $u=M/r^2$, $w=\sqrt{N}$,
 depends on $M$ (since
 $\beta=8\pi M$ for Schwarzschild background).
 This limiting configuration has also unit magnetic and
 electric charges. A detailed study of the solutions of (\ref{newSD}) 
 will be presented elsewhere. 

The situation is definitely different
when the solutions of the second branch are examined
in the same limit. Here the function $N(r)$ develops a local
maximum and a local  minimum respectively at  $r=r_{max}$ and $r=r_{min}$,
with        $r_h < r_{max} < r_{min}< \infty$.
The value $N(r_{min})$ got deeper and deeper for decreasing $\gamma$.
Our numerics  strongly suggests that the minimum of $N$  reaches the value zero
in the $\gamma=0$-limit and that the solution
approaches the self-dual solution   (\ref{newSD})
corresponding to $M = r_{min}/2$
on the interval $r\in [r_{min},\infty]$. At the same time the profiles
stay non trivial on $r\in [r_h, r_{min}]$.

%%%%%%%%%%%%%%%%%%%%%%%%%%%%%%%%%%%%%%%%%%%%%%%%%%%%%%%
\section{Euclidean action and thermodynamics}
%%%%%%%%%%%%%%%%%%%%%%%%%%%%%%%%%%%%%%%%%%%%%%%%%%%%%%%
Accordingly to Gibbons and Hawking \cite{Gibbons:1976ue}, thermodynamic functions including 
the entropy can be computed directly from 
the saddle point approximation to the gravitational partition function 
(namely the generating functional analytically 
continued to the Euclidean spacetime).
In the semiclassical approximation, the dominant 
 contribution to the path integral will 
come from the neighborhood of saddle points of the action, that is, of classical 
solution; the zeroth order contribution to 
$\log Z$ will be $-I$.

The Euclidean action of these solutions can be computed by using the
standard techniques in the literature.
Taking the trace of the metric equations of motion  
yields $R=8 \pi G (\partial_{a}\phi \partial^{a}\phi)$,
so the first two terms in the action cancel.
The dilaton equation of motion shows that the 
third term is a total derivative.
Thus the bulk action is proportional to the dilaton charge
$I_B=- \beta Q_d/{2\gamma}$,
and the total action of any solution can be recast as a boundary term
\footnote {In this Section we do not consider rescaled variables.}
(where $n^\mu$ is a unit outward pointing normal to the boundary)
\begin{eqnarray}
\label{rel1} 
I=-\frac{1}{8 \pi G}\int_{\partial\mathcal{M}} d^{3}x\sqrt{h}
e^{\phi/\gamma}\nabla_\mu (e^{-\phi/\gamma} n^{\mu}),
\end{eqnarray}
which diverges as $r \to \infty$.
In the traditional Euclidean path integral approach 
to black hole thermodynamics \cite{Gibbons:1976ue}, 
one has to choose a suitable reference background
and substract it in order to get a finite Euclidean action.
The reference background here is the flat four dimensional metric, although
there are some ambiguities related to the presence of gauge field charges
\cite{Hawking:1995ap}.
In a more recent approach, no reference background is required,
the action being regularized  
by adding suitable coordinate invariant  boundary 
surface counterterms to the gravitational action.
The suitable expression in our case is \cite{Kraus:1999di}
$I_{ct}=
\frac{1}{8 \pi G} \int d^3 x \sqrt{h} \sqrt{2\mathcal{R}},$
where $\mathcal{R}$ is the Ricci scalar of the boundary metric.
Varying the total action 
with respect to the
boundary metric $h_{\mu \nu}$, 
one computes also a divergence-free boundary stress-tensor.
The solutions' mass is 
the conserved charge associated with the Killing vector
$\partial/\partial \tau$ of the boundary metric
(see $e.g.$ \cite{Astefanesei:2005ad} and the more general approach in recent work
\cite{Mann:2005yr}).
By employing this approach or by subtracting the Minkowski 
background contribution, one finds
$I= \beta (M-Q_d/{ \gamma})/2.$
 
However, the dilaton charge is not an independent quantity, as proven by an alternative
computation of total action  (see also \cite{Kleihaus:2002tc} for the case
of a spacetime with Lorentzian signature). 
By integrating the Killing identity
$\nabla^a\nabla_b K_a=R_{bc}K^c,$
for the Killing field $K^a=\delta^a_\tau$, together with the Einstein equation
\begin{eqnarray}
\label{Rtt}
\frac{1}{8 \pi G}R_\tau^\tau 
-2e^{2\gamma \phi}{\rm Tr}(F_{\mu \tau}F^{\mu \tau})=
\frac{R}{16 \pi G}-\frac{1}{2}\partial_{a}\phi \partial^{a}\phi
-\frac{1}{2} e^{2\gamma \phi}{\rm Tr}(F_{ab}F^{ab})~
,
\end{eqnarray}
it is possible to isolate the bulk action contribution at infinity and at $r=0$ or $r=r_h$.
By using the YM equations and the asymptotic expansion (\ref{bc3}) we find
\begin{eqnarray}
\label{rel2} 
I=\beta(M-T\frac{A_H}{4G}-\Phi Q_e),
\end{eqnarray}
which implies
\begin{eqnarray}
\label{rel3} 
Q_d={\gamma}(M-2T\frac{A_H}{4G}-2 \Phi Q_e),
\end{eqnarray}
$A_H=4\pi r_h^2$ being the event horizon area
(with $A_H=0$ for topologically trivial solutions).
Here we should remark that
the variation of the action (\ref{lag0}) will give the correct equations of motion only if the 
 gauge potential $A_a$ is held fixed on the boundary $\partial M$.
This imposes the boundary condition $\delta A_{a}=0$ on $\partial M$.
Thus the action (\ref{lag0}) is  appropriate to study the ensemble with
fixed electric potential and fixed magnetic charge
\footnote{
To study the canonical enesemble with fixed magnetic {\emph and} electric charges,
we have to add a boundary term to impose fixed $Q_e$ 
as boundary condition at infinity \cite{Hawking:1995ap}.
The appropriate action in this case is 
$\tilde I=I-\frac{1}{4\pi G}\int_{\partial M} d^3 x \sqrt{h} n_a {\rm Tr}(F^{ab}A_b)$.
}.
This is the grand canonical ensemble, at fixed temperature and fixed potential.
The grand canonical (Gibbs) potential is $W=I/\beta=E-TS-\Phi Q_E$.

By using the approach in  \cite{Heusler:1993ke}  it can be proven that these solutions 
satisfy the first law of thermodynamics $dM=TdS-\Phi dQ_e$. 
As a result, one finds the thermodynamic quantities
\begin{eqnarray}
E= \left(\frac{\partial I}{\partial\beta}\right)_\Phi
-\frac{\Phi}{\beta}\left(\frac{\partial I}{\partial\Phi}\right)_\beta
=M, ~~
S= 
\beta\left(\frac{\partial I}{\partial\beta}\right)_\Phi-I
=\frac{A_H}{4G}~,
\label{grandc}
\end{eqnarray}
while the electric charge defined as $-\frac{1}{\beta}\left( {\partial
I}/{\partial\Phi}\right)_\beta$ is just $Q_e$.

As concerning the thermodynamic stability of these solutions,
for the parameters range we studied so far, we found that all
bolt solutions have a negative specific heat 
$C=T(\partial S/\partial T)_{\Phi}<0$.
 
%%%%%%%%%%%%%%%%%%%%%%%%%%%%%%%%%%%%%%%%%%%%%%%%%%%%%%%
\section{Further remarks}
%%%%%%%%%%%%%%%%%%%%%%%%%%%%%%%%%%%%%%%%%%%%%%%%%%%%%%%
In this paper we have investigated the basic properties
of a new type of dyonic solutions in Euclidean 
EYMd theory.
For an Euclidean spacetime signature, this results in
solution with a nonvanishing electric potential.
The existence of these configurations originates in the fact that, different from 
the abelian case, the asymptotic
magnitude of the nonabelian electric potential has a physical significance and
it cannot be gauged away without making the configurations time-dependent.
For the first type of configurations, the resulting manifolds have 
(if one identifies imaginary time) topology $R^3\times S^1$
and Euler number $\chi=0$  in 
contrast to the Euclidean black hole (-nonextremal) case  with $\chi=2$, 
thus possesing thermodynamic properties. 

It is obvious that these solutions have no reasonable Lorentzian counterparts. 
However, they possess a finite mass 
and action and give new saddle points of the EYMd Euclidean path integral.
Their basic properties are governed by the asymptotic magnitude of the electric potential
and the value of the dilaton coupling constant.

It is worthwhile to remark that
for a dilaton coupling constant $\gamma=1/\sqrt{3}$, these solutions 
can be uplifted to $d=5$, according to (with $z-$ the extra dimension)
\begin{eqnarray}
\label{uplift1}
ds^2_5 = e^{- \gamma\phi }
\left (\sigma^2(r) N(r) d\tau^2 + \frac{dr^2}{N(r)}  + r^2(d \theta^2 + \sin^2 \theta 
d \varphi^2)\right)+ e^{ 2\gamma\phi }dz ^2,
\end{eqnarray}
to become solutions in a five dimensional Euclidean EYM theory.
The  five  dimensional YM potential is still given by (\ref{YMansatz}).
Now one can perform a  Kaluza-Klein reduction to $d=4$ with respect the
Killing vector $\partial/\partial \tau$.
One finds in this way solutions in a four
dimensional EYMd-Higgs theory, the action principle
(\ref{lag0}) being  
supplemented by the Higgs term $e^{-2\gamma\bar{\phi}}
Tr(D_a\Phi D^a\Phi)$ \cite{Volkov:2001tb,Brihaye:2005pz},
$\bar{\phi}$ being the new $d=4$ dilaton field
\begin{eqnarray}
\label{dil4}
\bar{\phi}=-\frac{\phi}{2}+\frac{1}{2\gamma}\log N+\frac{1}{\gamma}\log \sigma.
\end{eqnarray}
The SU(2) field has a purely magnetic potential
\begin{eqnarray}
\label{reduced1}
A_a dx^a=\frac{1}{2g} \bigg\{    
 w(r) \tau_1  d \theta
+\left(\cot \theta \tau_3
+ w(r) \tau_2 \right) \sin \theta d \varphi \bigg \}, 
\end{eqnarray}
while the Higgs field is given by $\Phi=u(r)\tau_3/2 $.
Since the new  nonabelian field does not present an electric potential and $\partial/\partial z$
is still a Killing vector, these are solutions on both Lorentzian and Euclidean sections, with
\begin{eqnarray}
\label{uplift}
ds^2_4=e^{- 3\gamma \phi/2}\sqrt{N}\sigma(r)
 \left(\frac{dr^2}{N(r)}  + r^2(d \theta^2 + \sin^2 \theta d \varphi^2) 
 \pm e^{3\gamma\phi}  dz^2\right).
\end{eqnarray}
The solutions "dual" to topologically $R^4$ Euclidean configurations 
would describe globally regular monopole configurations,
originally studied in \cite{Volkov:2001tb},
and thus presenting a complicated branch structure,
with a critical value of $\alpha$.

Returning to the case
of a general $\gamma$, a discussion of possible generalizations of this work should start with
the radially excited solutions, for which the gauge function $\omega(r)$ possesses nodes.
We expect also the existence  axially symmetric solutions,
with a  nontrivial $\theta$-coordinate dependence.
The case of rotating generalizations
of the BK  solutions is particularly interesting, since
no spinning EYM solitons appear to exist on the Lorentzian section 
 \cite{VanderBij:2001nm,Kleihaus:2002ee}.
The nonexistence
of a  Lorentzian rotating generalization of the BK solution can be viewed as a
consequence of the impossibility to obtain regular, electrically
charged nonabelian solutions without a Higgs field.
This obstacle is avoided for an Euclidean signature and 
our preliminary numerical results indicate the existence
 of axially symmetric EYMd solitons with a nonzero
extradiagonal metric component $g_{\varphi \tau}$, 
usually associated with rotation for
a Lorentzian signature spacetime.
Since $\partial/\partial\varphi$ is still a Killing vector
for these solutions with $R^4$ topology, there is also a
new conserved quantity, $J$ (however, the
 magnetic charge is zero in this case). 
Similar "bolt" configurations generalizing for an Euclidean signature the 
rotating EYMd black holes found in \cite{Kleihaus:2003sh} 
are very likely to exist. 
\\
\\
{\bf\large Acknowledgements} \\
YB is grateful to the
Belgian FNRS for financial support.
The work of ER is carried out
in the framework of Enterprise--Ireland Basic Science Research Project
SC/2003/390 of Enterprise-Ireland.

 \end{document}